\documentclass[runningheads]{llncs}
\usepackage{hyperref}
\usepackage{caption}
\usepackage{colortbl}
\usepackage[table]{xcolor}

\usepackage{graphicx}
\begin{document}

\title{Fully Automated Tumor Segmentation for Brain MRI data using Multiplanner UNet.}

%
\author{Sumit Pandey\inst{1}\orcidID{2022-04-01 to 2025-03-31} \and
Satyasaran Changdar\inst{1}\orcidID{0000-0002-7704-8315} \and
 Mathias Perslev\inst{1}\orcidID{0000-0002-0358-4692}\and
Erik Bjørnager Dam\inst{1}\orcidID{0000-0002-8888-2524}} 
\authorrunning{s. Pandey et al.}
%
\institute{Department of Computer Science, Faculty of Science, University of Copenhagen, Denmark
\email{supa@di.ku.dk}\\
}
\maketitle              
\begin{abstract}
Automated segmentation of distinct tumor regions is critical for accurate diagnosis and treatment planning in pediatric brain tumors. This study evaluates the efficacy of the Multi-Planner U-Net (MPUnet) approach in segmenting different tumor subregions across three challenging datasets: Pediatrics Tumor Challenge (PED), Brain Metastasis Challenge (MET), and Sub-Sahara-Africa Adult Glioma (SSA). These datasets represent diverse scenarios and anatomical variations, making them suitable for assessing the robustness and generalization capabilities of the MPUnet model. By utilizing multi-planar information, the MPUnet architecture aims to enhance segmentation accuracy. Our results show varying performance levels across the evaluated challenges, with the tumor core (TC) class demonstrating relatively higher segmentation accuracy. However, variability is observed in the segmentation of other classes, such as the edema and enhancing tumor (ET) regions. These findings emphasize the complexity of brain tumor segmentation and highlight the potential for further refinement of the MPUnet approach and inclusion of MRI more data and preprocessing. 

\end{abstract}

\section{Introduction}
Automated tumor segmentation is crucial for surgical planning, treatment assessment, and long-term monitoring \cite{Boaro}. Manual segmentation is time-consuming and prone to variability. Pediatric brain tumors exhibit diverse characteristics, such as variable aggressiveness, prognosis, and heterogeneous histologic subregions \cite{Subramanian,Bakas}, making them challenging to assess. Pediatric tumors of the central nervous system, though rare, are the leading cause of cancer-related death in children. Unlike adult brain tumors, pediatric brain tumors have distinct imaging and clinical presentations. For instance, pediatric diffuse midline gliomas (DMGs), including the diffuse intrinsic pontine glioma (DIPG) subtype, are high-grade gliomas with a short average overall survival, similar to adult glioblastomas (GBMs) \cite{Kazerooni}. However, the incidence of GBMs is 3 in 100,000 people, while DMGs are about three times rarer. GBMs are typically found in the frontal or temporal lobes at an average age of 64 years, whereas DMGs are often located in the pons and diagnosed between 5 and 10 years of age \cite{Kazerooni,Davis}. Specific imaging tools are required for characterizing and diagnosing/prognosing pediatric brain tumors due to their unique features and challenges.

Some tumors, like DIPGs, are located in inaccessible areas and cannot be surgically removed, leading to reliance on size changes from longitudinal scans for assessing progression. The current standard uses 2D linear measurements, but these are inaccurate and increase inter-operator variability \cite{Bakas,Cooney}. Studies in adult brain tumors highlight the superiority of 3D volumetric measurements for predicting clinical outcomes. While volumetric tumor measurements are gaining recognition in assessing pediatric brain tumors, automated tools for segmenting tumor subregions are limited \cite{Kazerooni1}. Few methods exist, mainly focused on T2 FLAIR abnormal signal segmentation \cite{Nalepa}, but there's a lack of evaluation and comparison on the same data, leading to a gap in benchmarking automated segmentation tools for pediatric brain tumors.

Over the past 11 years, the MICCAI brain tumor segmentation (BraTS) challenges have created a benchmark dataset and community for adult glioma \cite{Menze,UBaid}. This year's BraTS has extended its focus to include a Cluster of Challenges \cite{BraTS-SSA_Adewole,BraTS-SSA_Adewole,BraTS-MET_Moawad,BraTS-MEN_LaBella,Karargyris,Bakas}, encompassing different tumor entities, missing data, and technical aspects. Notably, the BraTS 2022 challenge marked the first inclusion of pediatric brain tumors, particularly DMGs, in the test phase \cite{Kazerooni}. \\
Deep learning techniques have achieved impressive success across various domains of computer vision, including recognizing natural images, object detection, and image segmentation. This dominance extends to medical image segmentation \cite{Liang-Chieh} tasks as well. Initially introduced for biomedical image segmentation, the U-Net architecture has undergone several adaptations. For instance, a 3D version of U-Net was proposed, followed by the introduction of V-Net \cite{Fausto}, which incorporated residual blocks and the soft dice loss. Another enhancement involved incorporating attention \cite{Ozan} modules to reinforce the U-Net model. Beyond U-Net, researchers explored alternative architectures. Some studies segmented 3D volumes by slicing them into 2D sections and processing them using 2D segmentation networks. A hybrid approach \cite{Siqi} emerged with, which employed a 2D encoder based on ResNet50 and added 3D decoders. Furthermore, \cite{Yingda} fused 2D predictions through a 3D network to enhance predictions with contextual information. Nevertheless, U-Net derived architectures continue to maintain their supremacy in this domain. A recent advancement is the introduction of MPUnet by \cite{Perslev} , which secured the 5-th and 6-th in the first and second round of the 2018 Medical Segmentation Decathlon \cite{Antonell_2018_MPUNEt}. The system capitalizes on multi-planar data augmentation, enabling the utilization of a single 2D architecture inspired by the well-known U-Net design. Through multi-planar training, the system combines the efficiency of a 2D fully convolutional neural network with a structured augmentation strategy during both training and testing phases. This approach empowers the 2D model to acquire a comprehensive representation of the 3D image volume, thereby enhancing its ability to generalize effectively. For all the segmentation tasks in this study, we employed the MPUnet architecture.
\section{Methods}

\begin{figure}
\includegraphics[width=\textwidth]{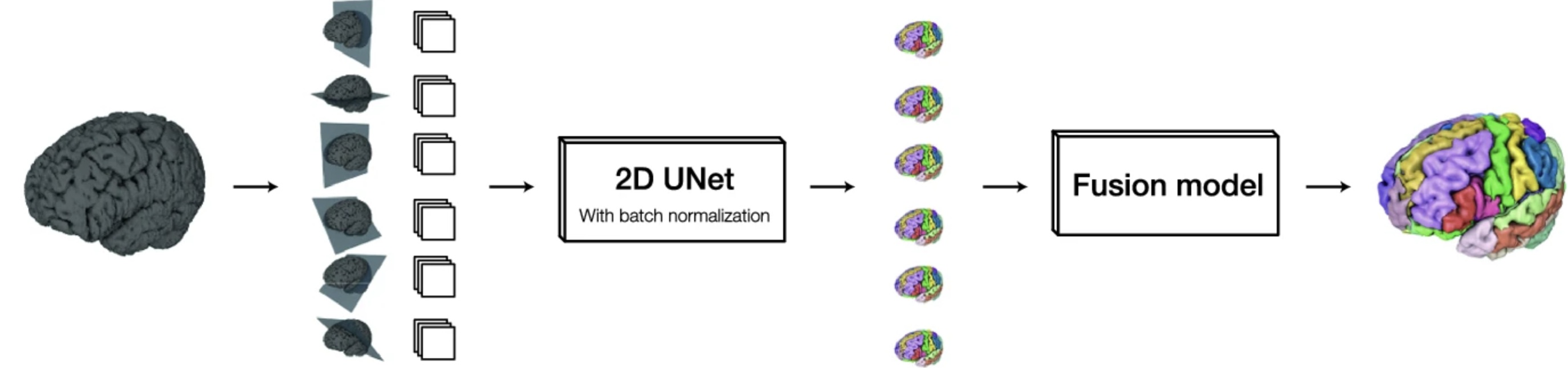}
\caption{The visual representation of the MPUnet model's schematic diagram \cite{Perslev}.} \label{fig1}
\end{figure}

The core methodology employed in this study centers around the Multi-Planner U-Net (MPUnet) architecture (see Fig.~\ref{fig1}), which serves as the backbone for our tumor segmentation approach. MPUnet is a modified variant of the U-Net architecture, which is a convolutional neural network designed for image segmentation tasks. MPUnet introduces a multi-planar training strategy, wherein the input image is rotated along various axes, generating multiple perspectives of the image. This technique enables the model to develop a comprehensive representation of the 3D image volume, thereby enhancing its ability to generalize and adapt to diverse anatomical variations and tumor appearances.

The evaluation was conducted using three distinct challenges: the Pediatrics Tumor Challenge (PED), the Brain Metastasis Challenge (MET), and the Sub-Sahara-Africa Adult Glioma Challenge (SSA). These challenges encompass a wide range of tumor types, anatomical variations, and imaging scenarios, providing a comprehensive assessment of the MPUnet's robustness and generalization capabilities.

To ensure reliable evaluation, a rigorous 3-fold cross-validation technique was employed. The dataset was randomly divided into three subsets: training, testing, and validation datasets. This division was repeated three times, effectively covering the dataset comprehensively while minimizing bias impact. Within each fold, two subsets were utilized for training the model, one for evaluating the model's performance, and the remaining subset for validation purposes. This approach enabled a thorough assessment of the model's generalization across different cases within the dataset.

\subsection{Preprocessing}
To address the limitations of GPU memory, we adopted a strategic approach by utilizing 
 only first three out of the four available imaging modalities for all three challenges: Pediatrics Tumor Challenge (PED), Brain Metastasis Challenge (MET), and Sub-Sahara-Africa Adult Glioma (SSA). This selective utilization of modalities aimed to optimize computational efficiency while ensuring a balanced trade-off between memory constraints and retaining essential information for accurate tumor segmentation.

\subsection{Statistical measurement}
To evaluate the precision of our generated masks, we utilized the Dice Score matrix as the primary evaluation metric across all three distinct classes: ET (enhancing tumor), TC (tumor core), and WT (whole tumor). Widely recognized in the field of image segmentation, the Dice Score quantifies the level of concurrence between the segmentation masks predicted by our model and the reference ground truth masks. This established metric enabled us to effectively measure the extent to which our segmentation results aligned with the actual tumor boundaries across the Pediatrics Tumor Challenge (PED), Brain Metastasis Challenge (MET), and Sub-Sahara-Africa Adult Glioma (SSA) datasets.

\subsection{Results}

The MPUNet performance results show mean and standard deviation values for Dice scores across three different challenges: PED, MET, and SSA on validation. The Dice scores are reported for three classes: ET, TC, and WT, representing different aspects of the segmentation quality.

\begin{table}
\centering
\caption{Dice Scores for Different Categories}
\setlength{\tabcolsep}{15pt}
\renewcommand{\arraystretch}{1.5}
\begin{tabular}{ |>{\columncolor{gray!20}}l|c|c|c| }
\hline

\rowcolor{gray}
\multicolumn{4}{|c|}{PED} \\
\hline
& Dice\_ET & Dice\_TC & Dice\_WT \\
\hline
Mean & 0.259 & 0.382 & 0.478 \\
Std & 0.377 & 0.370 & 0.370 \\
\hline
\rowcolor{gray}
\multicolumn{4}{|c|}{MET} \\
\hline
& Dice\_ET & Dice\_TC & Dice\_WT \\
\hline
Mean & 0.268 & 0.297 & 0.2832 \\
Std & 0.261 & 0.281 & 0.296 \\
\hline
\rowcolor{gray}
\multicolumn{4}{|c|}{SSA} \\
\hline
& Dice\_ET & Dice\_TC & Dice\_WT \\
\hline
Mean & 0.489 & 0.504 & 0.380 \\
Std & 0.3278 & 0.342 & 0.238 \\
\hline
\end{tabular}
\end{table}

In the PED challenge, the MPUNet achieved a mean Dice score of 0.259 for ET, 0.382 for TC, and 0.477 for WT. The standard deviations were 0.376 for ET, 0.37 for TC, and 0.369 for WT. These scores suggest that the model's performance varies across different classes, with TC having the highest mean score, indicating better segmentation results in that category. However, the relatively high standard deviations suggest some variability in the model's performance, particularly for ET and WT.

In the MET challenge, the MPUNet achieved a mean Dice score of 0.268 for ET, 0.297 for TC, and 0.283 for WT. The standard deviations were 0.261 for ET, 0.287 for TC, and 0.296 for WT. Here, the performance seems to be relatively balanced among the classes, with the lowest mean score being for WT. However, the standard deviations remain moderate, indicating some variability in performance.

In the SSA challenge, the MPUNet achieved a mean Dice score of 0.489 for ET, 0.504 for TC, and 0.38 for WT. The standard deviations were 0.327 for ET, 0.342 for TC, and 0.238 for WT. These results suggest that the model performed relatively well in the TC class, achieving the highest mean score among the challenges. The lower mean score for WT might indicate room for improvement in the segmentation of this particular class.

Overall, the MPUNet demonstrates varying levels of performance across different challenges and classes, with some challenges showing more consistent results than others. The relatively high standard deviations in some cases highlight the importance of robustness in the model's segmentation predictions, indicating the potential need for further optimization and fine-tuning to improve the overall segmentation accuracy, particularly in classes with lower mean Dice scores.

\section{Discussion}
The utilization of the MPUNet architecture for the segmentation of distinct tumor subregions across different datasets demonstrates the potential for enhanced accuracy in automated tumor segmentation. Our approach, which incorporates multi-planar information, addresses the complexity of pediatric brain tumor segmentation by aiming to capture the heterogeneity and anatomical variations present in these tumors.

The evaluation across the Pediatrics Tumor Challenge (PED), Brain Metastasis Challenge (MET), and Sub-Sahara-Africa Adult Glioma (SSA) datasets showcased varying levels of segmentation performance. The relatively higher accuracy observed in the tumor core (TC) class suggests the potential of the MPUnet approach in capturing the central regions of the tumors. However, the segmentation of other classes, such as the edema and enhancing tumor (ET) regions, exhibited more variability. This variability may stem from the unique characteristics and complexities associated with different tumor types and imaging modalities.

The complexity of pediatric brain tumors is further highlighted by the challenges in segmentation across diverse datasets. The inherent variability in tumor appearance, size, and location makes achieving consistent segmentation results a formidable task. Additionally, while the MPUnet architecture leverages multi-planar information to enhance segmentation, it may encounter difficulties in capturing the intricate details of certain tumor subregions, leading to variability in performance.

The results underscore the need for ongoing refinements and enhancements in the segmentation methodologies, especially for challenging cases such as pediatric brain tumors. Incorporating more comprehensive data, exploring advanced preprocessing techniques, and fine-tuning the MPUnet architecture for specific tumor types and subregions could potentially lead to improved segmentation accuracy and consistency.

\section{Conclusion}
In conclusion, the study's evaluation of the MPUNet approach for segmenting distinct tumor subregions in pediatric brain tumors across diverse datasets highlights its potential in enhancing accuracy, particularly in capturing tumor core regions. However, variability in segmenting other classes like edema and enhancing tumor regions underscores the intricate nature of pediatric brain tumor segmentation. The method's reliance on multi-planar information showcases its adaptability to anatomical variations and diverse tumor appearances. Challenges across Pediatrics Tumor Challenge (PED), Brain Metastasis Challenge (MET), and Sub-Sahara-Africa Adult Glioma (SSA) datasets underscore the need for continuous refinement, incorporating comprehensive data, advanced preprocessing, and targeted architecture enhancements to further improve segmentation accuracy and advance the field of automated tumor segmentation for pediatric brain tumors.

%
%
%

\begin{thebibliography}{8}

\bibitem{Boaro}
Boaro, Alessandro, Jakub R. Kaczmarzyk, Vasileios K. Kavouridis, Maya Harary, Marco Mammi, Hassan Dawood, Alice Shea et al.: Deep neural networks allow expert-level brain meningioma segmentation and present potential for improvement of clinical practice. Scientific Reports, 12, no. 1 15462 (2022)

\bibitem{Subramanian}
Subramanian S, Ahmad T.: Childhood Brain Tumors. In: StatPearls.:  StatPearls Publishing. Treasure Island (FL), PMID: 30571036 (2022)
\bibitem{Bakas}
Bakas, Spyridon, et al.: GLISTRboost: combining multimodal MRI segmentation, registration, and biophysical tumor growth modeling with gradient boosting machines for glioma segmentation. Brainlesion: Glioma, Multiple Sclerosis, Stroke and Traumatic Brain Injuries: First International Workshop, Brainles 2015, Held in Conjunction with MICCAI 2015, Munich, Germany, October 5, 2015, Revised Selected Papers 1. Springer International Publishing, (2016)
\bibitem{Kazerooni}
Kazerooni, Anahita Fathi, Nastaran Khalili, Xinyang Liu, Debanjan Haldar, Zhifan Jiang, Syed Muhammed Anwar, Jake Albrecht et al.: The Brain Tumor Segmentation (BraTS) Challenge 2023: Focus on Pediatrics (CBTN-CONNECT-DIPGR-ASNR-MICCAI BraTS-PEDs). arXiv preprint arXiv, 2305.17033 (2023)
\bibitem{Davis}
Davis ME. Glioblastoma.: Overview of Disease and Treatment. Clin J Oncol Nurs, Oct 1;20(5 Suppl):S2-8. doi: 10.1188/16.CJON.S1.2-8. PMID: 27668386; PMCID: PMC5123811 (2016)
\bibitem{Cooney}
T. M. Cooney, K. J. Cohen, C. V. Guimaraes, G. Dhall, J. Leach, M. Massimino, A. Erbetta, L. Chiapparini, F. Malbari, K. Kramer, et al.: Response assessment in diffuse intrinsic pontine glioma: recommendations from the response assessment in pediatric neuro-oncology (rapno) working group. The Lancet Oncology, vol. 21, no. 6, pp. e330–e336, (2020)
\bibitem{Kazerooni1}
A. Fathi Kazerooni, S. Arif, R. Madhogarhia, N. Khalili, D. Haldar, S. Bagheri, A. M. Familiar, H. Anderson, S. Haldar, W. Tu, et al.: Automated tumor segmentation and brain tissue extraction from multiparametric mri of pediatric brain tumors: A multi-institutional study, Neuro-Oncology Advances, p. vdad027, (2023)
\bibitem{Nalepa}
J. Nalepa, S. Adamski, K. Kotowski, S. Chelstowska, M. Machnikowska-Sokolowska, O. Bozek, A. Wisz, and E. Jurkiewicz.: Segmenting pediatric optic pathway gliomas from mri using deep learning,” Computers in Biology and Medicine, vol. 142, p. 105237, (2022)
\bibitem{Menze}
 B. H. Menze, A. Jakab, S. Bauer, J. Kalpathy-Cramer, K. Farahani, J. Kirby, Y. Burren, N. Porz, J. Slotboom, R. Wiest, et al.: The multimodal brain tumor image segmentation benchmark (brats). IEEE transactions on medical imaging, vol. 34, no. 10, pp. 1993–2024, (2014)
\bibitem{Antonell_2018_MPUNEt}
 Antonelli, M., Reinke, A., Bakas, S. et al.: The Medical Segmentation Decathlon. Nat Commun 13, 4128 (2022). https://doi.org/10.1038/s41467-022-30695-9
 
\bibitem{UBaid}
U. Baid, S. Ghodasara, S. Mohan, M. Bilello, E. Calabrese, E. Colak, K. Farahani, J. Kalpathy-Cramer, F. C. Kitamura, S. Pati, et al.: The rsna-asnr-miccai brats 2021 benchmark on brain tumor segmentation and radiogenomic classification, arXiv preprint arXiv:2107.02314, (2021)

\bibitem{BraTS-SSA_Adewole}
Adewole M, Rudie JD, Gbadamosi A, et al.: The Brain Tumor Segmentation (BraTS) Challenge 2023: Glioma Segmentation in Sub-Saharan Africa Patient Population (BraTS-Africa). arXiv:2305.19369 [eess.IV] (2023).
arXiv: https://arxiv.org/abs/2305.19369

\bibitem{BraTS-SSA_Adewole}
Kazerooni, A.F., Khalili, N., Liu, X., Haldar, D., Jiang, Z., Anwar, S.M., Albrecht, J., Adewole, M., Anazodo, U., Anderson, H. and Bagheri, S.: The Brain Tumor Segmentation (BraTS) Challenge 2023: Focus on Pediatrics (CBTN-CONNECT-DIPGR-ASNR-MICCAI BraTS-PEDs) (2023). arXiv preprint arXiv:2305.17033.


\bibitem{BraTS-MET_Moawad}

Moawad, A.W., Janas, A., Baid, U., Ramakrishnan, D., Jekel, L., Krantchev, K., Moy, H., Saluja, R., Osenberg, K., Wilms, K. and Kaur, M., 2023. The Brain Tumor Segmentation (BraTS-METS) Challenge 2023: Brain Metastasis Segmentation on Pre-treatment MRI (2023). arXiv preprint arXiv:2306.00838.


\bibitem{BraTS-MEN_LaBella}
LaBella, D., Adewole, M., Alonso-Basanta, M., Altes, T., Anwar, S.M., Baid, U., Bergquist, T., Bhalerao, R., Chen, S., Chung, V. and Conte, G.M., 2023. The ASNR-MICCAI Brain Tumor Segmentation (BraTS) Challenge 2023: Intracranial Meningioma (2023). arXiv preprint arXiv:2305.07642.


\bibitem{Karargyris}
Karargyris, A., Umeton, R., Sheller, M.J., Aristizabal, A., George, J., Wuest, A., Pati, S., Kassem, H., Zenk, M., Baid, U. and Narayana Moorthy, P.: Federated benchmarking of medical artificial intelligence with MedPerf. Nature Machine Intelligence, pp.1-12 (2023)
\bibitem{Bakas}
S. Bakas, H. Akbari, A. Sotiras, M. Bilello, M. Rozycki, J.S. Kirby, et al.: Advancing The Cancer Genome Atlas glioma MRI collections with expert segmentation labels and radiomic features, Nature Scientific Data, 4:170117 (2017) DOI: 10.1038/sdata.2017.117


\bibitem{Perslev}
Perslev, M., Dam, E.B., Pai, A., Igel, C.: One Network to Segment Them All: A General, Lightweight System for Accurate 3D Medical Image Segmentation. In: Shen, D., et al. Medical Image Computing and Computer Assisted Intervention – MICCAI 2019. MICCAI 2019. Lecture Notes in Computer Science(), vol 11765. Springer, Cham. (2019) https://doi.org/10.1007/978-3-030-32245-8
\bibitem{Yingda}
Yingda Xia, Lingxi Xie, Fengze Liu, Zhuotun Zhu, Elliot K Fishman, and Alan L Yuille.: Bridging the gap between 2d and 3d organ segmentation with volumetric fusion net. In MICCAI, pages 445–453. Springer, (2018) 
\bibitem{Liang-Chieh}
Liang-Chieh Chen, George Papandreou, Iasonas Kokkinos, Kevin Murphy, and Alan L Yuille. Deeplab: Semantic image segmentation with deep convolutional nets, atrous convolution, and fully connected crfs. PAMI, 40(4):834–848, (2018)
\bibitem{Fausto}
Fausto Milletari, Nassir Navab, and Seyed-Ahmad Ahmadi.: V-net: Fully convolutional neural networks for volumetric medical image segmentation In 3DV, pages 565–571. IEEE,(2016)
\bibitem{Ozan}
Ozan Oktay, Jo Schlemper, Loic Le Folgoc, Matthew Lee,
Mattias Heinrich, Kazunari Misawa, Kensaku Mori, Steven McDonagh, Nils Y Hammerla, Bernhard Kainz, et al.: Attention u-net: Learning where to look for the pancreas. MIDL, (2018)
\bibitem{Siqi}
Siqi Liu, Daguang Xu, S Kevin Zhou, Olivier Pauly, Sasa Grbic, Thomas Mertelmeier, Julia Wicklein, Anna Jerebko, Weidong Cai, and Dorin Comaniciu. 3d anisotropic hybrid network: Transferring convolutional features from 2d im- ages to 3d anisotropic volumes. In MICCAI, pages 851–858.
Springer, (2018)

\end{thebibliography}
%

\end{document}